\documentclass[aps,prl,reprint,showpacs,superscriptaddress,longbibliography]{revtex4-1}

\usepackage{graphicx} 
\usepackage{amsmath,amssymb}
\usepackage{textcomp}
\usepackage{color}

\usepackage{multirow}
\usepackage{makecell}

\usepackage[colorlinks=true,linkcolor=black,citecolor=black,urlcolor=black]{hyperref}

\begin{document}

\title{Real Topology on Symmetry-Invariant Loops Enforces Superconducting Pairing Zeros
}

\author{Hongyu Lu}
\affiliation{New Cornerstone Science Lab, Department of Physics, The University of Hong Kong, Hong Kong, China}
\affiliation{HK Institute of Quantum Science \& Technology, The University of Hong Kong, Hong Kong, China}
\affiliation{State Key Laboratory of Optical Quantum Materials, The University of Hong Kong, Hong Kong, China}

\author{Hao Jin}
\affiliation{New Cornerstone Science Lab, Department of Physics, The University of Hong Kong, Hong Kong, China}
\affiliation{HK Institute of Quantum Science \& Technology, The University of Hong Kong, Hong Kong, China}
\affiliation{State Key Laboratory of Optical Quantum Materials, The University of Hong Kong, Hong Kong, China}

\author{Wang Yao}
\email{wangyao@hku.hk}
\affiliation{New Cornerstone Science Lab, Department of Physics, The University of Hong Kong, Hong Kong, China}
\affiliation{HK Institute of Quantum Science \& Technology, The University of Hong Kong, Hong Kong, China}
\affiliation{State Key Laboratory of Optical Quantum Materials, The University of Hong Kong, Hong Kong, China}

\begin{abstract}
Normal-state topology can obstruct superconducting pairing and give rise to gapless nodes in the quasiparticle spectrum.
However, existing pairing-obstruction mechanisms largely rely on Fermi-surface topology or global Bloch-band topology.
Here, we establish a new and general mechanism: one-dimensional (1D) real topology, defined on antiunitary-symmetry-invariant loops, enforces pairing zeros without requiring Fermi-surface or global Bloch-band topology.
We derive sufficient conditions and lower bounds for pairing zeros across diverse pairing channels, for both zero and finite Cooper-pair momenta. 
The theorems apply to systems of arbitrary spatial dimensions
without requiring the pairing bands to be globally isolated.
We further show that such topologically enforced pairing zeros, if isolated, necessarily carry odd vorticity and can give rise to Bogoliubov–de Gennes nodes and zero-energy boundary modes, as demonstrated in a minimal two-orbital model. 
Our results reshape the understanding of how normal-state topology can obstruct the superconducting order,  highlight the significant role of 1D real topology beyond single-particle band phenomena, and provide a general framework for identifying candidate materials hosting nodal superconductivity.
\end{abstract}

\date{\today }
\maketitle

{{\it Introduction.}---}
The nodal structure of a superconductor plays a central role in determining the low-energy quasiparticle spectrum and  revealing the symmetry and microscopic nature of the pairing state~\cite{RevModPhys.72.969,Chen2014_review_ironbased}.
Conventionally, superconducting nodes appear when the symmetry of the pairing order parameter forces the gap function to vanish at certain  momenta~\cite{RevModPhys.63.239,Blount1985_symmetry_triplet_SC,Ono2022_symmetry_SC_nodes}.
Beyond such pairing-symmetry constraints, it has been recognized that normal-state topology can itself enforce zeros of the superconducting gap function. \cite{Murakami2003_Berry_phase_magnetic_SC,Li_Haldane2018_topological_nodal_cooper_pairing}.
However, to the best of our knowledge, existing mechanisms for topologically enforced nodal pairing rely on Fermi-surface or global Bloch-band topology~\cite{Murakami2003_Berry_phase_magnetic_SC,Li_Haldane2018_topological_nodal_cooper_pairing,sun2020_z2topologicallyobstructedsuperconducting, Lo2022_inherited_topological_sc_dirac, Euler_obstructed_Yu2022,Yu2023_Euler-obstructed, Wang2024_molecular_pairing_tbg, Zhu2025_Dipole-obstructed}.

Topological invariants, on the other hand, can also be defined on lower-dimensional submanifolds of momentum space, rather than on the Fermi surface or over the entire Brillouin zone (BZ). 
In particular, one-dimensional (1D) closed loops, invariant under suitable antiunitary symmetries, can carry nontrivial real topology~\cite{Ahn2019SW}.
While such topology has been extensively studied in band theory, its implications for Cooper pairing remain unexplored.

In this work, we establish a new mechanism of nodal pairing enforced by a nontrivial 1D real topology.
Crutially, this 1D real topology is defined on an antiunitary-symmetry-invariant submanifold of the BZ, rather than on the Fermi surface itself.
On these real loops, the nontrivial topology constrains the sign structure of the projected pairing and thus enforces pairing zeros.
Although the relevant topology is 1D, the pairing manifold, Fermi surface, and underlying Bloch system can all have arbitrary dimensions.
The mechanism applies to general pairing channels with either zero or finite Cooper-pair momentum. 
Furthermore, we show that such topologically enforced pairing zeros, when isolated, necessarily carry odd vorticity.
In both flat and dispersive Bloch bands, such pairing zeros can become Bogoliubov–de Gennes (BdG) nodes with inherited vorticity and give rise to localized zero-energy edge modes.
Importantly, our theorem holds in considerable generality: (i) it neither relies on nor precludes the global topology of the Fermi surface or Bloch bands; (ii) it requires no band isolation across the BZ, with the sole constraint that the relevant 1D loop exhibits no intra‑flavor band touching/crossing.

{{\it General pairing channels and normal-state 1D real topology.}---}
We start from a two-dimensional (2D) Bloch system with $N$ orbitals, and consider a general pairing  
between two Bloch bands labeled by
$f=1,2$, which can correspond to different fermion flavors or two distinct bands within the same flavor.
The general pairing Hamiltonian can be written as:
\begin{equation}
H_\Delta=
\frac{1}{2}\sum_{\mathbf{k},f,f'}
\Gamma_{ff'}\delta_\mathbf{Q}^{ff'}(\mathbf{k})\gamma_{\mathbf{k},f}^\dagger
\gamma_{\mathbf{Q}-\mathbf{k},f'}^\dagger
+\mathrm{h.c.},
\end{equation}
where $\gamma_{\mathbf{k},f}$ annihilates a fermion in band sector $f$, and
\begin{equation}
\delta_\mathbf{Q}^{ff'}(\mathbf{k})=
u^{(f)\dagger}(\mathbf{k}) D_\mathbf{Q}(\mathbf{k}) u^{(f')*}(\mathbf{Q}-\mathbf{k})
\label{define_pairing_function_kernel}
\end{equation}
is the pairing function in the band basis.
Here, $D_\mathbf{Q}(\mathbf{k})$ is the pairing kernel matrix in the orbital basis, and $\mathbf{Q}$ denotes
the Cooper-pair center-of-mass momentum, so that this formulation encompasses both zero-$\mathbf{Q}$ and finite-$\mathbf{Q}$ pairing.
We distinguish exchange-even and exchange-odd pairing channels according to the symmetry of the pairing kernel under simultaneous exchange of orbital indices and paired momenta:
\begin{equation}
D_\mathbf{Q}(\mathbf{k})=
\begin{cases}
+D_\mathbf{Q}^{T}(\mathbf{Q}-\mathbf{k}),&\mathrm{exchange-even},\\
-D_\mathbf{Q}^{T}(\mathbf{Q}-\mathbf{k}),&\mathrm{exchange-odd}.
\end{cases}
\end{equation}
For the two types of pairing channels, fermionic antisymmetry then fixes the band-sector-exchange structure described by the off-diagonal $2 \times 2$ matrix $\Gamma_{f,f'}$ - that is, how it transforms under $f \leftrightarrow f'$ - as follows:
\begin{equation}
\Gamma=
\begin{cases}
i\tau_y, & \mathrm{exchange-even},\\
\tau_x, & \mathrm{exchange-odd}.
\end{cases}
\end{equation}
Single-flavor intraband pairing is also considered, as discussed later.

\begin{table*}[htp!]
\centering
\renewcommand{\arraystretch}{1.4}
\setlength{\tabcolsep}{4pt}
\begin{tabular}{|c|c|c|c|c|c|}
\hline
Case & Topology & \makecell{Bloch wavefunctions \\ of the two paired sectors} & \makecell{Exchange parity\\ of the pairing channel}
    & Sufficient conditions & \makecell{Enforced zeros} \\
\hline

1 & \multirow{4}{*}{\raisebox{-8pt}{$\mathbb{Z}_2$}}
  & \multirow{2}{*}{Distinct}
  & Even
  & \multirow{2}{*}{$w_1^{(f)}+w_1^{(f')}=1\pmod2$}
  & \multirow{2}{*}{$N_\mathrm{zero}\ge1$} \\
\cline{1-1}\cline{4-4}

2 & & & Odd & & \\
\cline{1-1}\cline{3-6}

3 & & \multirow{2}{*}{\raisebox{-4pt}{Identical}} & Even
  & \makecell{$w_1=1$,\\ $D_\mathbf{Q}>0$ at $\mathbf{k_\ast}$}
  & $N_\mathrm{zero}\ge2$ \\
\cline{1-1}\cline{4-6}

4 & & & Odd
  & \multicolumn{2}{|c|}{\makecell{No topology-enforced zeros;\\
antisymmetry-enforced zeros exist.}} \\
\hline

5 & \multirow{4}{*}{$\mathbb{Z}$}
  & Distinct
  & \multirow{2}{*}{Even}
  & \multirow{2}{*}{\makecell{$|w^{(f)}+w^{(f')}|>0$\\$D_\mathbf{Q}>0$ along $\mathcal{L}_f$}}
  & \multirow{2}{*}{$N_\mathrm{zero}\ge|w^{(f)}+w^{(f')}|$} \\
\cline{1-1}\cline{3-3}

6 & & Identical & & & \\
\cline{1-1}\cline{3-6}

7 & & Distinct
  & \multirow{2}{*}{Odd}
  & \multirow{2}{*}{\makecell{$|w^{(f)}+w^{(f')}|>0$\\$D_\mathbf{Q}^T=-D_{\mathbf{Q}}$ along $\mathcal{L}_f$}}
  & \multirow{2}{*}{$N_\mathrm{zero}\ge|w^{(f)}+w^{(f')}|$} \\
\cline{1-1}\cline{3-3}

8 & & Identical & & & \\
\hline

\end{tabular}
\caption{\textbf{Summary of topologically enforced pairing zeros by 1D real topology.}
We consider eight cases based on three criteria: (i) whether the 1D topology is characterized by an integer $\mathbb{Z}$ winding number $w$ or only a $\mathbb{Z}_2$ invariant $w_1$; (ii) whether the two paired sectors have identical or distinct Bloch wavefunctions; and (iii) the exchange parity of the pairing channel.
$D_\mathbf{Q}$ is the pairing kernel in the orbital basis, as defined in Eq.~\ref{define_pairing_function_kernel}, and $N_\mathrm{zero}$ denotes the number of pairing zeros on $\mathcal{L}_f$.
$k_\ast$ refers to the two pairing-invariant momenta that remain unchanged under the exchange $\mathbf{k}\leftrightarrow\mathbf{Q}-\mathbf{k}\pmod{\mathbf{G}}$.
In addition, single-flavor intraband pairing within a single Bloch band can be treated similarly to cases 4 and 8.}
\label{tab_theorem_classification}
\end{table*}


We consider a 2D Bloch system possessing an antiunitary symmetry $\mathcal{A}$ belonging to the class of $\mathcal{A}^2=1$, and with at least one closed loop $\mathcal{L}$ in the BZ that is pointwise invariant under $\mathcal{A}$. 
Along $\mathcal{L}$, the normalized Bloch state $|u^{(f)}_\mathbf{k}\rangle$ is then allowed to be real, which is defined only up to an overall sign.
Therefore, it naturally takes values in the real projective space $\mathbb{RP}^{N-1}$.
For systems with $N=2$ orbitals, $\mathbb{RP}^{1}\simeq S^{1}$, and the Bloch states along $\mathcal{L}$ are characterized by an integer winding number:
\begin{equation}
    w^{(f)}=\frac{1}{\pi}\oint_{\mathcal{L}} d\theta^{(f)}(\mathbf{k}),
\ \mathrm{for}\ 
|u^{(f)}_{\mathbf {k}}\rangle=
\begin{pmatrix}
\cos\theta^{(f)}(\mathbf{k}),\
\sin\theta^{(f)}(\mathbf{k})
\end{pmatrix},
\label{eq_definition_w}
\end{equation}
with $w^{(f)}\in\mathbb Z$.
For $N\ge3$, the additional orbital dimensions allow even windings to be continuously unwound, leaving only a $\mathbb{Z}_2$ classification, which is characterized by the first Stiefel–Whitney number $w_1$~\cite{Ahn2019SW}, defined as:\begin{equation}
|u^{(f)}_{\mathbf k_f}\rangle=(-1)^{w^{(f)}_1}|u^{(f)}_{\mathbf{k}_i}\rangle,\quad w_1^{(f)}\in\mathbb{Z}_2.
\label{eq_definition_w1}
\end{equation}
Here $\mathbf{k}_i$ and $\mathbf{k}_f$ denote the same point on $\mathcal{L}$, with $|u^{(f)}_{\mathbf{k}_f} \rangle$ being the state obtained by parallel transporting $|u^{(f)}_{\mathbf{k}_i} \rangle $ once around the loop $\mathcal{L}$.
A nontrivial $w^{(f)}_1=1$ indicates an obstruction to a globally continuous and periodic real gauge. 
Note that, the formulation 
applies equally to lattice and continuum Bloch systems.
We also emphasize that although the relevant topology is 1D, the pairing function is defined over the full 2D BZ.

{{\it Theorems of topologically enforced pairing zeros.}---}
The obstruction by the underlying 1D real topology applies broadly to different pairing channels, while the sufficient conditions for the enforced pairing zeros may differ. 
As summarized in Tab.~\ref{tab_theorem_classification}, we organize our analysis into eight cases based on three criteria: (i) whether the 1D topology is characterized by an integer $\mathbb{Z}$ winding number or only a $\mathbb{Z}_2$ invariant; (ii) whether the two paired band sectors $f$ and $f'$ have identical or distinct Bloch wavefunctions; and (iii) the exchange parity of the pairing channel.

\begin{figure}[htp!]
	\centering		
	\includegraphics[width=0.5\textwidth]{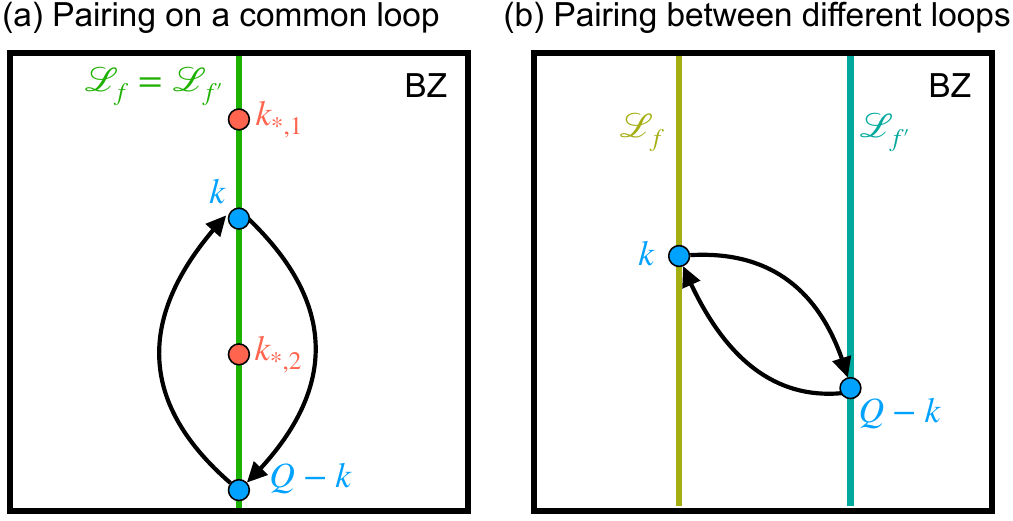}
	\caption{\textbf{Schematics of Cooper pairing on symmetry-invariant loops.} (a) Pairing on a common loop, $\mathcal{L}_f=\mathcal{L}_{f'}$, with two pairing-invariant  points $\mathbf{k}_{\ast,1}$ and $\mathbf k_{\ast,2}$ satisfying $2\mathbf{k}_\ast=\mathbf Q$ (mod $\mathbf{G}$). 
    (b) Pairing between different loops, $\mathcal{L}_f\neq\mathcal{L}_{f'}$, where $\mathbf{k}\in\mathcal{L}_f$ pairs with $\mathbf{Q}-\mathbf{k}\in\mathcal{L}_{f'}$.
    Both geometries accommodate zero- and finite-$\mathbf Q$ pairing, with identical or distinct Bloch wavefunctions as classified in Tab.~\ref{tab_theorem_classification}.
    }
	\label{fig_pairing_map_schematic}
\end{figure}

For all cases, we consider Cooper-pair momenta $\mathbf{Q}$ that can map the two loops into each other:
\begin{equation}
\mathbf{k}\in\mathcal{L}_f
\quad\Rightarrow\quad
\mathbf{Q}-\mathbf{k}\in\mathcal{L}_{f'}
\pmod{\mathbf{G}},
\end{equation}
where $\mathbf G$ is a reciprocal lattice vector. 
Whether $\mathcal{L}_f$ and $\mathcal{L}_{f'}$ are identical or distinct does not affect our analysis.
The schematics of pairing  either on a common symmetry-invariant loop, or between different loops, are shown in Fig.~\ref{fig_pairing_map_schematic}.
We further assume that pairing function $\delta_{\mathbf Q}^{ff'}(\mathbf k)$ can be chosen real along the loops, as naturally realized by pairing that preserves the antiunitary symmetry.
Below we analyze the eight cases in detail.

In cases 1 and 2 where the Bloch wavefunctions of the two paired sectors are distinct, and the 1D topology on $\mathcal{L}_{f}$ and $\mathcal{L}_{f'}$ is characterized by a $\mathbb{Z}_2$ invariant,
the sufficient condition for having enforced pairing zeros is:
\begin{equation}w_1^{(f)}+w_1^{(f')}=1\pmod2.\label{contion_case1&2}\end{equation}
This simply follows from the fact that the pairing function in continuous real gauges acquires the following sign after winding once around the paired loops:
\begin{equation}
\delta_\mathbf{Q}^{ff'}
\rightarrow
(-1)^{w_1^{(f)}+w_1^{(f')}}
\delta_\mathbf{Q}^{ff'}.
\end{equation}
The condition in Eq.~\ref{contion_case1&2} then enforces the real pairing amplitude to change sign after one circuit, leading to at least one zero of $\delta^{ff'}_\mathbf{Q}$ on $\mathcal{L}_f$.

Next we turn to cases where the two paired bands have identical Bloch wavefunctions and 1D topology along a common $\mathcal{L}$ is still characterized by a $\mathbb Z_2$ invariant. The general condition in Eq.~\ref{contion_case1&2} and the associated result does not apply in such cases, as
$w_1^{(f)}+w_1^{(f')}=0\pmod 2$. 
Nevertheless, for exchange-even pairing (case 3 in Table~\ref{tab_theorem_classification}), having $w_1^{(f)}=w_1^{(f')}=1$ can still enforce at least two zeros if the pairing kernel $D_{\mathbf Q}$ is positive-definite at the two pairing-invariant momenta $\mathbf k_{\ast, i}$ $(i=1,2)$ on $\mathcal{L}$, i.e. $\mathbf k_{\ast,i}=\mathbf Q-\mathbf k_{\ast,i}\pmod{\mathbf G}$ (as shown in Fig.~\ref{fig_pairing_map_schematic}(a)).
Positive definiteness of $D_{\mathbf{Q}}$ is defined in the orbital basis where the Bloch states are real.
To prove this, we note that $\mathbf k_{\ast, 1}$ and $\mathbf k_{\ast, 2}$ divide $\mathcal{L}$ into two arcs $\mathcal{L}_\pm$ that are related by the pairing map $\mathbf{k}\leftrightarrow\mathbf{Q}-\mathbf{k}$.
The real pairing functions on the two arcs $\mathcal{L}_\pm$ are given by
\begin{equation}\delta_\mathbf{Q}^{\pm}(\mathbf{k})
=u_\pm^T(\mathbf{k})D_{\mathbf Q}(\mathbf{k})
u_\mp(\mathbf{Q}-\mathbf{k}).
\end{equation}
Continuous real gauges on the two arcs satisfy
\begin{equation}
u_-(\mathbf{k}_{\ast,i})
=\eta_i u_+(\mathbf{k}_{\ast,i}),
\quad \eta_{i=1}=-\eta_{i=2},
\end{equation}
where the opposite signs follow from $w_1=1$.
Having $D_{\mathbf Q}(\mathbf{k}_{\ast,i}) >0$, the signs of $\delta_\mathbf{Q}^{\pm}$ at $\mathbf{k}_{\ast,i}$ are determined by $\eta_i$ and are thus opposite.
By continuity, each arc must contain at least one zero, thus enforcing at least two zeros along $\mathcal{L}$.

In contrast, for exchange-odd pairing between two bands of identical Bloch wavefunctions (case 4), the nontrivial 1D $\mathbb Z_2$ topology does not enforce zeros. Instead, irrespective of the 1D topology, the exchange antisymmetry enforces zeros right at two pairing-invariant momenta:
\begin{equation}
    \delta_\mathbf{Q}(\mathbf{k}_{\ast,i})
=-\delta_\mathbf{Q}(\mathbf{Q}-\mathbf{k}_{\ast,i})=-\delta_\mathbf{Q}(\mathbf{k}_{\ast,i})=0.
\end{equation}

In cases 5-8, the 1D topology along the paired loops is characterized by integer windings $w^{(f)}, w^{(f')}$, allowing stronger lower bounds of topologically enforced zeros.
Regardless of whether the two band sectors have identical or distinct Bloch wavefunctions and whether the pairing is exchange-even or exchange-odd, at least $|w^{(f)}+w^{(f')}|$ zeros are topologically enforced along $\mathcal L_f$ under the following sufficient conditions.
For exchange-even pairing, the sufficient condition is the positive-definiteness of the pairing kernel $D_\mathbf{Q}$ along $\mathcal{L}_f$, which ensures that the pairing function takes the form:
\begin{equation}
\delta_\mathbf{Q}^{ff'}(\mathbf{k})
\propto\cos[\theta_f(\mathbf{k})-\tilde\theta_{f'}(\mathbf{Q}-\mathbf{k})].
\end{equation}
Here $\tilde\theta_{f'}$ denotes the angle of $|u^{(f')}\rangle$ smoothly deformed by the positive-definite $D_{\mathbf Q}$ without changing its winding number.
For exchange-odd pairing, the sufficient condition is $D_\mathbf{Q}^T=-D_{\mathbf{Q}}$ along $\mathcal{L}_f$, which dictates an alternative form of the pairing function:
\begin{equation}
\delta_\mathbf{Q}^{ff'}(\mathbf{k})
\propto\sin[\theta_{f'}(\mathbf{Q}-\mathbf{k})-\theta_f(\mathbf{k})].
\label{delta_Z_odd_sin}
\end{equation}
Under either pairing exchange parities, the relative angle winds by $(w^{(f)}+w^{(f')})\pi$, enforcing at least $|w^{(f)}+w^{(f')}|$ zeros on $\mathcal{L}_f$.
In addition, in case 8 with exchange-odd pairing between identical Bloch wavefunctions, exchange antisymmetry also enforces zeros at the pairing-invariant momenta $k_\ast$, irrespective of the winding number.

While the above eight cases are for pairing channels with definite exchange parity, mixtures of exchange-even and exchange-odd pairing are also possible. 
The 1D $\mathbb{Z}_2$ topology can still enforce pairing zeros without additional conditions, as the results established above apply directly to the mixed pairing. The
1D $\mathbb{Z}$ topology can also enforce zeros under the mixture, provided suitable conditions are met by the combined pairing kernel; we do not further elaborate on this latter case here.

Moreover, single-flavor intraband pairing within a single Bloch band can be treated similarly to case 8 (for $\mathbb{Z}$ topology) and case 4 (for $\mathbb{Z}_2$ topology). 
Under the same pairing-kernel conditions, the $\mathbb{Z}$ case yields $N_\mathrm{zero}\ge2|w|$, while the $\mathbb{Z}_2$ case guarantees only the antisymmetry-enforced zeros at the pairing-invariant momenta.

We emphasize again that while the pairing function is defined over the entire BZ, our analysis of sufficient conditions for topologically enforced zeros has so far been based only on the properties of the 1D loops.
The general theorems do not require the pairing Bloch bands to be globally isolated from other bands; nondegeneracy is required only along the corresponding real loops within each flavor sector.

{{\it Vorticity of isolated pairing zeros.}---}
We further show that an isolated pairing zero enforced by the above theorems necessarily carries an odd vorticity.
Consider such a zero at $\mathbf{k}_0\in\mathcal{L}$. 
We define the vorticity on a sufficiently small loop $\mathcal{C}$ enclosing $\mathbf{k}_0$ as:
\begin{equation}
    \mathcal{V}(\mathbf{k}_0)=\frac{1}{2\pi}
\oint_{\mathcal{C}}d\mathbf{k}\cdot\nabla_\mathbf{k}
\arg\delta_\mathbf{Q}(\mathbf{k}).
\end{equation}
The loop $\mathcal{C}$ crosses $\mathcal{L}$ at two points on opposite sides of $\mathbf{k}_0$. Since the pairing function $\delta_\mathbf{Q}$ is real on $\mathcal{L}$ and changes sign across the zero, its phases at these two points differ by $\pi$ (mod $2\pi$). 
The antiunitary symmetry relates the two halves of $\mathcal C$ by complex conjugation, ensuring an odd phase winding:
\begin{equation}
   \mathcal{V}(\mathbf{k}_0)\in 2\mathbb{Z}+1.
\end{equation}
Notably, such pairing vortices can arise independently of the global topology of the underlying Bloch bands or Fermi surface. 
Nontrivial Chern topology imposes an additional constraint on the total vorticity of the pairing zeros~\cite{Murakami2003_Berry_phase_magnetic_SC,Li_Haldane2018_topological_nodal_cooper_pairing}, but is not required for their existence.

{{\it BdG nodes.}---}
We next discuss how these topologically enforced pairing zeros manifest as nodal quasiparticles in the superconducting states.
In the projected band, the BdG Hamiltonian for a pairing state with momentum $\mathbf{Q}$ can be written as:
\begin{equation}
H_{\mathrm{BdG}}(\mathbf{k})=
\begin{pmatrix}
\xi_{\mathbf{k}} & \delta_{\mathbf{Q}}(\mathbf{k})\\
\delta_{\mathbf{Q}}^\ast(\mathbf{k}) & -\xi_{\mathbf{Q}-\mathbf{k}}
\end{pmatrix},
\end{equation}
where $\xi_\mathbf{k}=\epsilon_\mathbf{k}-\mu$, with $\epsilon_\mathbf{k}$ and $\mu$ being the band dispersion and chemical potential.
Then the quasiparticle energies are:
$E_{\pm}(\mathbf{k})=
\frac{\xi_\mathbf{k}-\xi_{\mathbf{Q}-\mathbf{k}}}{2}
\pm\sqrt{(\frac{\xi_\mathbf{k}+\xi_{\mathbf{Q}-\mathbf{k}}}{2})^2
+|\delta_\mathbf{Q}(\mathbf{k})|^2 }$.
For an exactly flat band tuned to the Fermi level, 
$\xi_\mathbf{k}=0$, therefore every topologically enforced pairing zero directly becomes a BdG node carrying the same vorticity as that pairing zero. 
For a dispersive band, an enforced pairing zero becomes a BdG node when either $\xi_\mathbf{k}=0$ or $\xi_{\mathbf{Q}-\mathbf{k}}=0$.
Therefore, nodal superconductivity can be enforced by a  1D real topology of the normal-state Bloch wave functions, even when neither the Fermi surface nor the pairing Bloch bands carry any nontrivial global topology.

\begin{figure}[htp!]
	\centering		
	\includegraphics[width=0.5\textwidth]{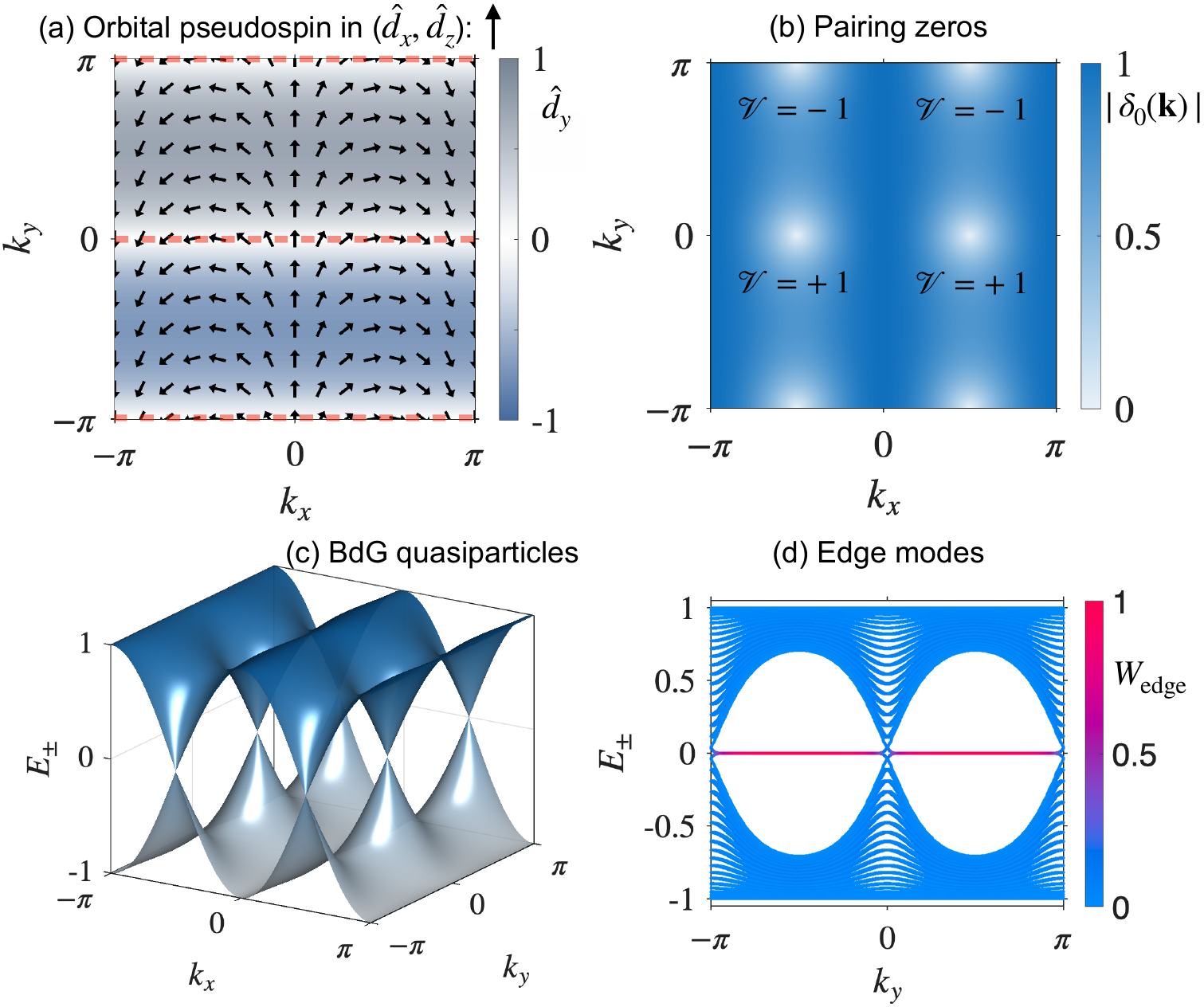}
	\caption{\textbf{A minimal example.} (a) The pseudospin texture of the minimal two-orbital model. The colormap refers to the $\hat{d}_y$ component and the black arrows refer to the (x)-(z) projection.
    There are two real loops $\mathcal{L}_{k_y=0,\pi}$ indicated by the red dashed lines, both of which have an $w=1$ invariant.
    (b) Projected pairing function of the onsite $\mathbf{Q}=0$ flavor-singlet channel. There are four enforced zeros with $\mathcal{V}=\pm1$ vortices.
    (c) The BdG quasiparticle spectrum with resulting nodes and inherited vorticity, at half filling of the lower flat band.
    (d) Localized zero-energy edge modes for open boundary conditions along $x$ direction. The edge weight $W_\mathrm{edge}$ denotes the total BdG probability on the two outermost boundaries.
    }
	\label{fig_minimal_model}
\end{figure}

{{\it Demonstration with a minimal two-orbital model.}---}
For concrete demonstrations, we consider a minimal two-orbital model:
\begin{equation}
H_0=
\sum_{\mathbf k,\alpha}
c_{\mathbf{k}\alpha}^{\dagger}
[\hat{\mathbf{d}}(\mathbf{k})\cdot\boldsymbol{\sigma}]
c_{\mathbf{k}\alpha},
\quad \hat{\mathbf{d}}(\mathbf{k})
=\frac{\mathbf{d}(\mathbf{k})}{|\mathbf{d}(\mathbf{k})|},
\end{equation}
with $\mathbf d(\mathbf k)=(\sin{k_x},\sin{k_y},\cos{k_x})$. 
Here $\boldsymbol{\sigma}$ acts in the two-orbital space and $\alpha$ labels the two identical fermion flavors.
There are two exactly flat bands at energies $\epsilon_\pm=\pm1$ with trivial $C=0$ Chern numbers. 
Because the model lacks a global real structure, the 2D Euler class cannot be defined. 
Nonetheless, 
there is an antiunitary symmetry $\mathcal{A}=M_xT$  with $\mathcal{A}^2=1$ and  $H_0(k_x,-k_y)=H_0^\ast(k_x,k_y)$, giving two invariant real lines $\mathcal{L}_{k_y=0,\pi}$. 
The invariant loops both carry $\mathbb{Z}$ winding number $|w=1|$, with
the orbital pseudospin texture shown in Fig.~\ref{fig_minimal_model} (a).

We consider the onsite, zero-momentum ($\mathbf Q=0$) flavor-singlet pairing state,  with $D_{\mathbf{Q}=0}=I_2$ and the overall pairing amplitude set to unity for simplicity.
This realizes case 6 of our classification and satisfies its sufficient conditions.
The projected pairing function is shown in Fig.~\ref{fig_minimal_model}(b), with two zeros on each loop. 
Each zero on $\mathcal{L}_{k_y=0}$ carries a nontrivial vortex charge $\mathcal{V}=1$, while each zero on $\mathcal{L}_{k_y=\pi}$ carries $\mathcal{V}=-1$.
At half filling of the lower flat band, the BdG quasiparticle dispersions, $E_\pm(\mathbf{k})=\pm|\delta_0(\mathbf{k})|=\pm \sqrt{\frac{\cos^2 k_x+\sin^2 k_y}{1+\sin^2 k_y}}$, inherit the four nodes with corresponding vorticity on $\mathcal{L}_{k_y=0,\pi}$, as shown in Fig.~\ref{fig_minimal_model}(c).

When the resulting BdG nodes inherit nontrivial vortex charges from the pairing zeros, the nodal bulk-boundary correspondence  gives rise to zero-energy flat bands within momentum regions bounded by the projections of the bulk nodes onto the edge Brillouin zone~\cite{Sato2011Andreev,Schnyder2011FlatBands}.
This is demonstrated here by calculations with the open boundary conditions along $x$ direction, shown in Fig.~\ref{fig_minimal_model}(d).

{{\it Diverse antiunitary symmetries and arbitrary spatial dimensions.}---}
Although the relevant real topology is 1D, the theorems on the existence and the lower bounds of the pairing nodes apply to systems of arbitrary spatial dimensions. 
Moreover, different antiunitary symmetries at different system dimensionalities may lead to different structures of pairing zeros.

In 2D systems, $M_{x,y}T$ or $C_{2x,2y}T$ can define 1D real loops, where the enforced sign-changing zeros are generically isolated and carry odd vorticity. 
By contrast, $C_{2z}T$ or $PT$ can define a real structure throughout the BZ. 
The two independent noncontractible loops can then carry $(w_x,w_y)\in\mathbb Z^2$ or $(w_{1x},w_{1y})\in\mathbb Z_2^2$~\cite{Ahn2019SW}. 
The enforced zeros generically form nodal lines without vortex charges.

In 1D Bloch systems, $MT$, $C_2T$, or $PT$ symmetry can make the entire BZ real, giving rise to isolated zeros without momentum-space vorticity. 

In 3D systems, $M_iT$ and $C_{2i}T$ can define 1D real lines and 2D real planes, respectively. 
Since a complex pairing zero imposes two real conditions, the enforced zeros generically stretch into 1D lines. 
These pairing-zero lines generically intersect the real lines at isolated points, while those enforced on real planes lie within these planes.
In both cases, the topologically enforced pairing-zero lines generically carry odd transverse vorticity.
When $PT$ symmetry 
provides a real structure throughout the 3D BZ, 
the pairing zeros generically form 2D surfaces without vortex charges.

{{\it Discussion and outlook.}---}
The mechanism based on 1D real topology is fundamentally different from the usual topological classification of BdG quasiparticles~\cite{Sato2006_nodal_sc, Beri2010_topologically_stable, Schnyder2011FlatBands}, which protects nodes that already exist. In contrast, the 1D real topology of the normal state directly enforces the pairing zeros.
Beyond the minimal model presented for demonstration, the theorem itself is broadly applicable. 
It will be interesting to apply this mechanism in realistic materials of different dimensions, since the normal-state setting can be naturally realized in various platforms as discussed above, which would be valuable for identifying, predicting, and designing topologically enforced nodal superconductivity.
An interesting future direction is to explore whether 1D multiband real topology, including potential non-Abelian structures~\cite{Wu2019NonAbelian}, can enforce zeros in matrix-valued pairing involving more than two active bands.

Lastly, we note the relevant normal-state antiunitary symmetries ($\mathcal{A}^2=1$) can be realized in a broad range of crystalline and moir\'e systems. 
Examples include $C_{2z}T$-symmetric low-energy models of twisted bilayer graphene~\cite{Song2019_MATBG, Ahn2019_fragile_topology_tbg}, and twisted TMD homobilayers retaining in-plane $C_{2i}T$ symmetries~\cite{Xu2024MoTe2Wannier, Zhai2025TMDReview}.
Three-dimensional magnetic crystals, including kagome metals such as $\mathrm{Co_3Sn_2S_2}$~\cite{Schilberth2025Co3Sn2S2}, and altermagnets such as MnTe~\cite{Takahashi2025MnTe}, can support $M_iT$ symmetries.
These platforms motivate material-specific searches for nontrivial 1D real topology and pairing channels satisfying our criteria.

\begin{acknowledgments}
{{\it Acknowledgments}---}
The work is supported by the National Natural Science Foundation of China (No.12425406), Research Grant Council of Hong Kong (AoE/P-701/20, HKU SRFS21227S05), and New Cornerstone Science Foundation.
\end{acknowledgments}

\bibliographystyle{apsrev4-2}

\begin{thebibliography}{25}%
	\makeatletter
	\providecommand \@ifxundefined [1]{%
		\@ifx{#1\undefined}
	}%
	\providecommand \@ifnum [1]{%
		\ifnum #1\expandafter \@firstoftwo
		\else \expandafter \@secondoftwo
		\fi
	}%
	\providecommand \@ifx [1]{%
		\ifx #1\expandafter \@firstoftwo
		\else \expandafter \@secondoftwo
		\fi
	}%
	\providecommand \natexlab [1]{#1}%
	\providecommand \enquote  [1]{``#1''}%
	\providecommand \bibnamefont  [1]{#1}%
	\providecommand \bibfnamefont [1]{#1}%
	\providecommand \citenamefont [1]{#1}%
	\providecommand \href@noop [0]{\@secondoftwo}%
	\providecommand \href [0]{\begingroup \@sanitize@url \@href}%
	\providecommand \@href[1]{\@@startlink{#1}\@@href}%
	\providecommand \@@href[1]{\endgroup#1\@@endlink}%
	\providecommand \@sanitize@url [0]{\catcode `\\12\catcode `\$12\catcode
		`\&12\catcode `\#12\catcode `\^12\catcode `\_12\catcode `\%12\relax}%
	\providecommand \@@startlink[1]{}%
	\providecommand \@@endlink[0]{}%
	\providecommand \url  [0]{\begingroup\@sanitize@url \@url }%
	\providecommand \@url [1]{\endgroup\@href {#1}{\urlprefix }}%
	\providecommand \urlprefix  [0]{URL }%
	\providecommand \Eprint [0]{\href }%
	\providecommand \doibase [0]{https://doi.org/}%
	\providecommand \selectlanguage [0]{\@gobble}%
	\providecommand \bibinfo  [0]{\@secondoftwo}%
	\providecommand \bibfield  [0]{\@secondoftwo}%
	\providecommand \translation [1]{[#1]}%
	\providecommand \BibitemOpen [0]{}%
	\providecommand \bibitemStop [0]{}%
	\providecommand \bibitemNoStop [0]{.\EOS\space}%
	\providecommand \EOS [0]{\spacefactor3000\relax}%
	\providecommand \BibitemShut  [1]{\csname bibitem#1\endcsname}%
	\let\auto@bib@innerbib\@empty
	\bibitem [{\citenamefont {Tsuei}\ and\ \citenamefont
		{Kirtley}(2000)}]{RevModPhys.72.969}%
	\BibitemOpen
	\bibfield  {author} {\bibinfo {author} {\bibfnamefont {C.~C.}\ \bibnamefont
			{Tsuei}}\ and\ \bibinfo {author} {\bibfnamefont {J.~R.}\ \bibnamefont
			{Kirtley}},\ }\href {https://doi.org/10.1103/RevModPhys.72.969} {\bibfield
		{journal} {\bibinfo  {journal} {Rev. Mod. Phys.}\ }\textbf {\bibinfo {volume}
			{72}},\ \bibinfo {pages} {969} (\bibinfo {year} {2000})}\BibitemShut
	{NoStop}%
	\bibitem [{\citenamefont {Chen}\ \emph {et~al.}(2014)\citenamefont {Chen},
		\citenamefont {Dai}, \citenamefont {Feng}, \citenamefont {Xiang},\ and\
		\citenamefont {Zhang}}]{Chen2014_review_ironbased}%
	\BibitemOpen
	\bibfield  {author} {\bibinfo {author} {\bibfnamefont {X.}~\bibnamefont
			{Chen}}, \bibinfo {author} {\bibfnamefont {P.}~\bibnamefont {Dai}}, \bibinfo
		{author} {\bibfnamefont {D.}~\bibnamefont {Feng}}, \bibinfo {author}
		{\bibfnamefont {T.}~\bibnamefont {Xiang}},\ and\ \bibinfo {author}
		{\bibfnamefont {F.-C.}\ \bibnamefont {Zhang}},\ }\href
	{https://doi.org/10.1093/nsr/nwu007} {\bibfield  {journal} {\bibinfo
			{journal} {National Science Review}\ }\textbf {\bibinfo {volume} {1}},\
		\bibinfo {pages} {371} (\bibinfo {year} {2014})},\ \Eprint
	{https://arxiv.org/abs/https://academic.oup.com/nsr/article-pdf/1/3/371/31567405/nwu007.pdf}
	{https://academic.oup.com/nsr/article-pdf/1/3/371/31567405/nwu007.pdf}
	\BibitemShut {NoStop}%
	\bibitem [{\citenamefont {Sigrist}\ and\ \citenamefont
		{Ueda}(1991)}]{RevModPhys.63.239}%
	\BibitemOpen
	\bibfield  {author} {\bibinfo {author} {\bibfnamefont {M.}~\bibnamefont
			{Sigrist}}\ and\ \bibinfo {author} {\bibfnamefont {K.}~\bibnamefont {Ueda}},\
	}\href {https://doi.org/10.1103/RevModPhys.63.239} {\bibfield  {journal}
		{\bibinfo  {journal} {Rev. Mod. Phys.}\ }\textbf {\bibinfo {volume} {63}},\
		\bibinfo {pages} {239} (\bibinfo {year} {1991})}\BibitemShut {NoStop}%
	\bibitem [{\citenamefont {Blount}(1985)}]{Blount1985_symmetry_triplet_SC}%
	\BibitemOpen
	\bibfield  {author} {\bibinfo {author} {\bibfnamefont {E.~I.}\ \bibnamefont
			{Blount}},\ }\href {https://doi.org/10.1103/PhysRevB.32.2935} {\bibfield
		{journal} {\bibinfo  {journal} {Phys. Rev. B}\ }\textbf {\bibinfo {volume}
			{32}},\ \bibinfo {pages} {2935} (\bibinfo {year} {1985})}\BibitemShut
	{NoStop}%
	\bibitem [{\citenamefont {Ono}\ and\ \citenamefont
		{Shiozaki}(2022)}]{Ono2022_symmetry_SC_nodes}%
	\BibitemOpen
	\bibfield  {author} {\bibinfo {author} {\bibfnamefont {S.}~\bibnamefont
			{Ono}}\ and\ \bibinfo {author} {\bibfnamefont {K.}~\bibnamefont {Shiozaki}},\
	}\href {https://doi.org/10.1103/PhysRevX.12.011021} {\bibfield  {journal}
		{\bibinfo  {journal} {Phys. Rev. X}\ }\textbf {\bibinfo {volume} {12}},\
		\bibinfo {pages} {011021} (\bibinfo {year} {2022})}\BibitemShut {NoStop}%
	\bibitem [{\citenamefont {Murakami}\ and\ \citenamefont
		{Nagaosa}(2003)}]{Murakami2003_Berry_phase_magnetic_SC}%
	\BibitemOpen
	\bibfield  {author} {\bibinfo {author} {\bibfnamefont {S.}~\bibnamefont
			{Murakami}}\ and\ \bibinfo {author} {\bibfnamefont {N.}~\bibnamefont
			{Nagaosa}},\ }\href {https://doi.org/10.1103/PhysRevLett.90.057002}
	{\bibfield  {journal} {\bibinfo  {journal} {Phys. Rev. Lett.}\ }\textbf
		{\bibinfo {volume} {90}},\ \bibinfo {pages} {057002} (\bibinfo {year}
		{2003})}\BibitemShut {NoStop}%
	\bibitem [{\citenamefont {Li}\ and\ \citenamefont
		{Haldane}(2018)}]{Li_Haldane2018_topological_nodal_cooper_pairing}%
	\BibitemOpen
	\bibfield  {author} {\bibinfo {author} {\bibfnamefont {Y.}~\bibnamefont
			{Li}}\ and\ \bibinfo {author} {\bibfnamefont {F.~D.~M.}\ \bibnamefont
			{Haldane}},\ }\href {https://doi.org/10.1103/PhysRevLett.120.067003}
	{\bibfield  {journal} {\bibinfo  {journal} {Phys. Rev. Lett.}\ }\textbf
		{\bibinfo {volume} {120}},\ \bibinfo {pages} {067003} (\bibinfo {year}
		{2018})}\BibitemShut {NoStop}%
	\bibitem [{\citenamefont {Sun}\ and\ \citenamefont
		{Li}(2020)}]{sun2020_z2topologicallyobstructedsuperconducting}%
	\BibitemOpen
	\bibfield  {author} {\bibinfo {author} {\bibfnamefont {C.}~\bibnamefont
			{Sun}}\ and\ \bibinfo {author} {\bibfnamefont {Y.}~\bibnamefont {Li}},\
	}\href {https://arxiv.org/abs/2009.07263} {\bibinfo {title} {$\mathbb{Z}_2$
			topologically obstructed superconducting order}} (\bibinfo {year} {2020}),\
	\Eprint {https://arxiv.org/abs/2009.07263} {arXiv:2009.07263
		[cond-mat.supr-con]} \BibitemShut {NoStop}%
	\bibitem [{\citenamefont {Lo}\ \emph {et~al.}(2022)\citenamefont {Lo},
		\citenamefont {Po},\ and\ \citenamefont
		{Nevidomskyy}}]{Lo2022_inherited_topological_sc_dirac}%
	\BibitemOpen
	\bibfield  {author} {\bibinfo {author} {\bibfnamefont {C.~F.~B.}\
			\bibnamefont {Lo}}, \bibinfo {author} {\bibfnamefont {H.~C.}\ \bibnamefont
			{Po}},\ and\ \bibinfo {author} {\bibfnamefont {A.~H.}\ \bibnamefont
			{Nevidomskyy}},\ }\href {https://doi.org/10.1103/PhysRevB.105.104501}
	{\bibfield  {journal} {\bibinfo  {journal} {Phys. Rev. B}\ }\textbf {\bibinfo
			{volume} {105}},\ \bibinfo {pages} {104501} (\bibinfo {year}
		{2022})}\BibitemShut {NoStop}%
	\bibitem [{\citenamefont {Yu}\ \emph {et~al.}(2022)\citenamefont {Yu},
		\citenamefont {Chen},\ and\ \citenamefont
		{Das~Sarma}}]{Euler_obstructed_Yu2022}%
	\BibitemOpen
	\bibfield  {author} {\bibinfo {author} {\bibfnamefont {J.}~\bibnamefont
			{Yu}}, \bibinfo {author} {\bibfnamefont {Y.-A.}\ \bibnamefont {Chen}},\ and\
		\bibinfo {author} {\bibfnamefont {S.}~\bibnamefont {Das~Sarma}},\ }\href
	{https://doi.org/10.1103/PhysRevB.105.104515} {\bibfield  {journal} {\bibinfo
			{journal} {Phys. Rev. B}\ }\textbf {\bibinfo {volume} {105}},\ \bibinfo
		{pages} {104515} (\bibinfo {year} {2022})}\BibitemShut {NoStop}%
	\bibitem [{\citenamefont {Yu}\ \emph {et~al.}(2023)\citenamefont {Yu},
		\citenamefont {Xie}, \citenamefont {Wu},\ and\ \citenamefont
		{Das~Sarma}}]{Yu2023_Euler-obstructed}%
	\BibitemOpen
	\bibfield  {author} {\bibinfo {author} {\bibfnamefont {J.}~\bibnamefont
			{Yu}}, \bibinfo {author} {\bibfnamefont {M.}~\bibnamefont {Xie}}, \bibinfo
		{author} {\bibfnamefont {F.}~\bibnamefont {Wu}},\ and\ \bibinfo {author}
		{\bibfnamefont {S.}~\bibnamefont {Das~Sarma}},\ }\href
	{https://doi.org/10.1103/PhysRevB.107.L201106} {\bibfield  {journal}
		{\bibinfo  {journal} {Phys. Rev. B}\ }\textbf {\bibinfo {volume} {107}},\
		\bibinfo {pages} {L201106} (\bibinfo {year} {2023})}\BibitemShut {NoStop}%
	\bibitem [{\citenamefont {Wang}\ \emph {et~al.}(2024)\citenamefont {Wang},
		\citenamefont {Zhou}, \citenamefont {Peng}, \citenamefont {Lian},\ and\
		\citenamefont {Song}}]{Wang2024_molecular_pairing_tbg}%
	\BibitemOpen
	\bibfield  {author} {\bibinfo {author} {\bibfnamefont {Y.-J.}\ \bibnamefont
			{Wang}}, \bibinfo {author} {\bibfnamefont {G.-D.}\ \bibnamefont {Zhou}},
		\bibinfo {author} {\bibfnamefont {S.-Y.}\ \bibnamefont {Peng}}, \bibinfo
		{author} {\bibfnamefont {B.}~\bibnamefont {Lian}},\ and\ \bibinfo {author}
		{\bibfnamefont {Z.-D.}\ \bibnamefont {Song}},\ }\href
	{https://doi.org/10.1103/PhysRevLett.133.146001} {\bibfield  {journal}
		{\bibinfo  {journal} {Phys. Rev. Lett.}\ }\textbf {\bibinfo {volume} {133}},\
		\bibinfo {pages} {146001} (\bibinfo {year} {2024})}\BibitemShut {NoStop}%
	\bibitem [{\citenamefont {Zhu}\ and\ \citenamefont
		{Zhang}(2025)}]{Zhu2025_Dipole-obstructed}%
	\BibitemOpen
	\bibfield  {author} {\bibinfo {author} {\bibfnamefont {P.}~\bibnamefont
			{Zhu}}\ and\ \bibinfo {author} {\bibfnamefont {R.-X.}\ \bibnamefont
			{Zhang}},\ }\href {https://doi.org/10.1103/PhysRevB.111.L081107} {\bibfield
		{journal} {\bibinfo  {journal} {Phys. Rev. B}\ }\textbf {\bibinfo {volume}
			{111}},\ \bibinfo {pages} {L081107} (\bibinfo {year} {2025})}\BibitemShut
	{NoStop}%
	\bibitem [{\citenamefont {Ahn}\ \emph {et~al.}(2019{\natexlab{a}})\citenamefont
		{Ahn}, \citenamefont {Park}, \citenamefont {Kim}, \citenamefont {Kim},\ and\
		\citenamefont {Yang}}]{Ahn2019SW}%
	\BibitemOpen
	\bibfield  {author} {\bibinfo {author} {\bibfnamefont {J.}~\bibnamefont
			{Ahn}}, \bibinfo {author} {\bibfnamefont {S.}~\bibnamefont {Park}}, \bibinfo
		{author} {\bibfnamefont {D.}~\bibnamefont {Kim}}, \bibinfo {author}
		{\bibfnamefont {Y.}~\bibnamefont {Kim}},\ and\ \bibinfo {author}
		{\bibfnamefont {B.-J.}\ \bibnamefont {Yang}},\ }\href
	{https://doi.org/10.1088/1674-1056/ab4d3b} {\bibfield  {journal} {\bibinfo
			{journal} {Chinese Physics B}\ }\textbf {\bibinfo {volume} {28}},\ \bibinfo
		{pages} {117101} (\bibinfo {year} {2019}{\natexlab{a}})}\BibitemShut
	{NoStop}%
	\bibitem [{\citenamefont {Sato}\ \emph {et~al.}(2011)\citenamefont {Sato},
		\citenamefont {Tanaka}, \citenamefont {Yada},\ and\ \citenamefont
		{Yokoyama}}]{Sato2011Andreev}%
	\BibitemOpen
	\bibfield  {author} {\bibinfo {author} {\bibfnamefont {M.}~\bibnamefont
			{Sato}}, \bibinfo {author} {\bibfnamefont {Y.}~\bibnamefont {Tanaka}},
		\bibinfo {author} {\bibfnamefont {K.}~\bibnamefont {Yada}},\ and\ \bibinfo
		{author} {\bibfnamefont {T.}~\bibnamefont {Yokoyama}},\ }\href
	{https://doi.org/10.1103/PhysRevB.83.224511} {\bibfield  {journal} {\bibinfo
			{journal} {Phys. Rev. B}\ }\textbf {\bibinfo {volume} {83}},\ \bibinfo
		{pages} {224511} (\bibinfo {year} {2011})}\BibitemShut {NoStop}%
	\bibitem [{\citenamefont {Schnyder}\ and\ \citenamefont
		{Ryu}(2011)}]{Schnyder2011FlatBands}%
	\BibitemOpen
	\bibfield  {author} {\bibinfo {author} {\bibfnamefont {A.~P.}\ \bibnamefont
			{Schnyder}}\ and\ \bibinfo {author} {\bibfnamefont {S.}~\bibnamefont {Ryu}},\
	}\href {https://doi.org/10.1103/PhysRevB.84.060504} {\bibfield  {journal}
		{\bibinfo  {journal} {Phys. Rev. B}\ }\textbf {\bibinfo {volume} {84}},\
		\bibinfo {pages} {060504} (\bibinfo {year} {2011})}\BibitemShut {NoStop}%
	\bibitem [{\citenamefont {Sato}(2006)}]{Sato2006_nodal_sc}%
	\BibitemOpen
	\bibfield  {author} {\bibinfo {author} {\bibfnamefont {M.}~\bibnamefont
			{Sato}},\ }\href {https://doi.org/10.1103/PhysRevB.73.214502} {\bibfield
		{journal} {\bibinfo  {journal} {Phys. Rev. B}\ }\textbf {\bibinfo {volume}
			{73}},\ \bibinfo {pages} {214502} (\bibinfo {year} {2006})}\BibitemShut
	{NoStop}%
	\bibitem [{\citenamefont {B{\'e}ri}(2010)}]{Beri2010_topologically_stable}%
	\BibitemOpen
	\bibfield  {author} {\bibinfo {author} {\bibfnamefont {B.}~\bibnamefont
			{B{\'e}ri}},\ }\href {https://doi.org/10.1103/PhysRevB.81.134515} {\bibfield
		{journal} {\bibinfo  {journal} {Phys. Rev. B}\ }\textbf {\bibinfo {volume}
			{81}},\ \bibinfo {pages} {134515} (\bibinfo {year} {2010})}\BibitemShut
	{NoStop}%
	\bibitem [{\citenamefont {Wu}\ \emph {et~al.}(2019)\citenamefont {Wu},
		\citenamefont {Soluyanov},\ and\ \citenamefont
		{Bzdu{\v{s}}ek}}]{Wu2019NonAbelian}%
	\BibitemOpen
	\bibfield  {author} {\bibinfo {author} {\bibfnamefont {Q.}~\bibnamefont
			{Wu}}, \bibinfo {author} {\bibfnamefont {A.~A.}\ \bibnamefont {Soluyanov}},\
		and\ \bibinfo {author} {\bibfnamefont {T.}~\bibnamefont {Bzdu{\v{s}}ek}},\
	}\href {https://doi.org/10.1126/science.aau8740} {\bibfield  {journal}
		{\bibinfo  {journal} {Science}\ }\textbf {\bibinfo {volume} {365}},\ \bibinfo
		{pages} {1273} (\bibinfo {year} {2019})}\BibitemShut {NoStop}%
	\bibitem [{\citenamefont {Song}\ \emph {et~al.}(2019)\citenamefont {Song},
		\citenamefont {Wang}, \citenamefont {Shi}, \citenamefont {Li}, \citenamefont
		{Fang},\ and\ \citenamefont {Bernevig}}]{Song2019_MATBG}%
	\BibitemOpen
	\bibfield  {author} {\bibinfo {author} {\bibfnamefont {Z.}~\bibnamefont
			{Song}}, \bibinfo {author} {\bibfnamefont {Z.}~\bibnamefont {Wang}}, \bibinfo
		{author} {\bibfnamefont {W.}~\bibnamefont {Shi}}, \bibinfo {author}
		{\bibfnamefont {G.}~\bibnamefont {Li}}, \bibinfo {author} {\bibfnamefont
			{C.}~\bibnamefont {Fang}},\ and\ \bibinfo {author} {\bibfnamefont {B.~A.}\
			\bibnamefont {Bernevig}},\ }\href
	{https://doi.org/10.1103/PhysRevLett.123.036401} {\bibfield  {journal}
		{\bibinfo  {journal} {Phys. Rev. Lett.}\ }\textbf {\bibinfo {volume} {123}},\
		\bibinfo {pages} {036401} (\bibinfo {year} {2019})}\BibitemShut {NoStop}%
	\bibitem [{\citenamefont {Ahn}\ \emph {et~al.}(2019{\natexlab{b}})\citenamefont
		{Ahn}, \citenamefont {Park},\ and\ \citenamefont
		{Yang}}]{Ahn2019_fragile_topology_tbg}%
	\BibitemOpen
	\bibfield  {author} {\bibinfo {author} {\bibfnamefont {J.}~\bibnamefont
			{Ahn}}, \bibinfo {author} {\bibfnamefont {S.}~\bibnamefont {Park}},\ and\
		\bibinfo {author} {\bibfnamefont {B.-J.}\ \bibnamefont {Yang}},\ }\href
	{https://doi.org/10.1103/PhysRevX.9.021013} {\bibfield  {journal} {\bibinfo
			{journal} {Phys. Rev. X}\ }\textbf {\bibinfo {volume} {9}},\ \bibinfo {pages}
		{021013} (\bibinfo {year} {2019}{\natexlab{b}})}\BibitemShut {NoStop}%
	\bibitem [{\citenamefont {Xu}\ \emph {et~al.}(2024)\citenamefont {Xu},
		\citenamefont {Li}, \citenamefont {Xu}, \citenamefont {Bi},\ and\
		\citenamefont {Zhang}}]{Xu2024MoTe2Wannier}%
	\BibitemOpen
	\bibfield  {author} {\bibinfo {author} {\bibfnamefont {C.}~\bibnamefont
			{Xu}}, \bibinfo {author} {\bibfnamefont {J.}~\bibnamefont {Li}}, \bibinfo
		{author} {\bibfnamefont {Y.}~\bibnamefont {Xu}}, \bibinfo {author}
		{\bibfnamefont {Z.}~\bibnamefont {Bi}},\ and\ \bibinfo {author}
		{\bibfnamefont {Y.}~\bibnamefont {Zhang}},\ }\href
	{https://doi.org/10.1073/pnas.2316749121} {\bibfield  {journal} {\bibinfo
			{journal} {Proc. Natl. Acad. Sci. U.S.A.}\ }\textbf {\bibinfo {volume}
			{121}},\ \bibinfo {pages} {e2316749121} (\bibinfo {year} {2024})}\BibitemShut
	{NoStop}%
	\bibitem [{\citenamefont {Zhai}\ \emph {et~al.}(2025)\citenamefont {Zhai},
		\citenamefont {Yu},\ and\ \citenamefont {Yao}}]{Zhai2025TMDReview}%
	\BibitemOpen
	\bibfield  {author} {\bibinfo {author} {\bibfnamefont {D.}~\bibnamefont
			{Zhai}}, \bibinfo {author} {\bibfnamefont {H.}~\bibnamefont {Yu}},\ and\
		\bibinfo {author} {\bibfnamefont {W.}~\bibnamefont {Yao}},\ }\bibfield
	{journal} {\bibinfo  {journal} {Rep. Prog. Phys.}\ }\textbf {\bibinfo
		{volume} {88}},\ \href {https://doi.org/10.1088/1361-6633/adefef}
	{10.1088/1361-6633/adefef} (\bibinfo {year} {2025})\BibitemShut {NoStop}%
	\bibitem [{\citenamefont {Schilberth}\ \emph {et~al.}(2025)\citenamefont
		{Schilberth}, \citenamefont {Jiang}, \citenamefont {Le~Mardel{\'e}},
		\citenamefont {Papp}, \citenamefont {Mohelsky}, \citenamefont {Kassem},
		\citenamefont {Tabata}, \citenamefont {Waki}, \citenamefont {Nakamura},
		\citenamefont {Guo}, \citenamefont {Orlita}, \citenamefont {Arita},
		\citenamefont {K{\'e}zsm{\'a}rki},\ and\ \citenamefont
		{Bord{\'a}cs}}]{Schilberth2025Co3Sn2S2}%
	\BibitemOpen
	\bibfield  {author} {\bibinfo {author} {\bibfnamefont {F.}~\bibnamefont
			{Schilberth}}, \bibinfo {author} {\bibfnamefont {M.-C.}\ \bibnamefont
			{Jiang}}, \bibinfo {author} {\bibfnamefont {F.}~\bibnamefont
			{Le~Mardel{\'e}}}, \bibinfo {author} {\bibfnamefont {L.~B.}\ \bibnamefont
			{Papp}}, \bibinfo {author} {\bibfnamefont {I.}~\bibnamefont {Mohelsky}},
		\bibinfo {author} {\bibfnamefont {M.~A.}\ \bibnamefont {Kassem}}, \bibinfo
		{author} {\bibfnamefont {Y.}~\bibnamefont {Tabata}}, \bibinfo {author}
		{\bibfnamefont {T.}~\bibnamefont {Waki}}, \bibinfo {author} {\bibfnamefont
			{H.}~\bibnamefont {Nakamura}}, \bibinfo {author} {\bibfnamefont {G.-Y.}\
			\bibnamefont {Guo}}, \bibinfo {author} {\bibfnamefont {M.}~\bibnamefont
			{Orlita}}, \bibinfo {author} {\bibfnamefont {R.}~\bibnamefont {Arita}},
		\bibinfo {author} {\bibfnamefont {I.}~\bibnamefont {K{\'e}zsm{\'a}rki}},\
		and\ \bibinfo {author} {\bibfnamefont {S.}~\bibnamefont {Bord{\'a}cs}},\
	}\href {https://doi.org/10.1038/s41535-025-00785-0} {\bibfield  {journal}
		{\bibinfo  {journal} {npj Quantum Mater.}\ }\textbf {\bibinfo {volume}
			{10}},\ \bibinfo {pages} {67} (\bibinfo {year} {2025})}\BibitemShut {NoStop}%
	\bibitem [{\citenamefont {Takahashi}\ \emph {et~al.}(2025)\citenamefont
		{Takahashi}, \citenamefont {Huang}, \citenamefont {Yu} \emph
		{et~al.}}]{Takahashi2025MnTe}%
	\BibitemOpen
	\bibfield  {author} {\bibinfo {author} {\bibfnamefont {K.}~\bibnamefont
			{Takahashi}}, \bibinfo {author} {\bibfnamefont {H.-F.}\ \bibnamefont
			{Huang}}, \bibinfo {author} {\bibfnamefont {J.-X.}\ \bibnamefont {Yu}}, \emph
		{et~al.},\ }\href {https://doi.org/10.1038/s41535-025-00784-1} {\bibfield
		{journal} {\bibinfo  {journal} {npj Quantum Mater.}\ }\textbf {\bibinfo
			{volume} {10}},\ \bibinfo {pages} {70} (\bibinfo {year} {2025})}\BibitemShut
	{NoStop}%
\end{thebibliography}

\end{document}